\documentclass{article}
\usepackage[utf8x]{inputenc}
\usepackage{amsmath}
\usepackage{amssymb}
\usepackage{graphicx}
\usepackage{tabularx}
\newcommand{\beq}{\begin{equation}}
\newcommand{\eeq}{\end{equation}}
\newcommand{\beqn}{\begin{eqnarray}}
\newcommand{\eeqn}{\end{eqnarray}}
\newcommand{\bea}[1]{\beq\begin{array}{#1}}
\newcommand{\eea}{\end{array}\eeq}
\newcommand{\eq}[1]{(\ref{#1})}

\title{Lay-up Optimization of Laminated Composites:\\
Mixed Approach with Exact Feasibility Bounds on Lamination Parameters}
\author{
F.~Gubarev, V.~Kunin, A.~Pospelov \\[3mm]
DATADVANCE \\[2mm]
Pokrovsky blvd. 3 building 1B, 109028, Moscow, Russia
}
\date{}
\begin{document}
\maketitle
\begin{abstract}\noindent
We suggest modified bi-level approach for finding the best stacking sequence of laminated
composite structures subject to mechanical, blending and manufacturing constraints.
We propose to use both the number of plies laid up at predefined angles
and lamination parameters as independent variables at outer (global) stage of bi-level
scheme aimed to satisfy buckling, strain and percentage constraints.
Our formulation allows precise definition of the feasible region
of lamination parameters and greatly facilitates the solution of inner level
problem of finding the optimal stacking sequence.
\end{abstract}
\subsubsection*{Introduction}
Due to the excellent mechanical properties laminated composite materials are widely used nowadays in
various industries (see, e.g., Ref.~\cite{review} for review).
The common problem is to develop composite structures of minimal weight subject to
mechanical, blending and manufacturing (technological) constraints, the degrees of freedom being
the number of plies laid-up at a particular orientation (angle). Therefore, the industry faces
with extremely difficult high dimensional mixed-integer optimization problem, in which 
both total plies percentage and the concrete stacking sequence of plies are to be found in order to get
optimal design satisfying all the imposed constraints.

It is important to note that substantial part of constraint functions are
stacking sequence independent. Indeed, the weight of laminated composite
structure is simply proportional to the total number of plies.
On the other hand, imposed constraints could be naturally divided into two classes:
{\it i})``universal'' mechanical loads written in terms of various stiffness tensors;
{\it ii}) ``non-universal'' manufacturing constraints, which vary for different applications and dictate,
for instance, particular laminate blending rules (see Ref.~\cite{altair} for details and further examples).
Let us consider the most important mechanical constraints, which
according to classical laminate plate theory \cite{TsaiHahn} (for review see, e.g., Ref~\cite{Bettebghor}
and references therein) are formulated in terms of in-plane $A$, coupling $B$ and out-of-plane $D$ stiffness tensors
\beq
\label{mechanical_basics}
\left[ \begin{array}{c} N \\ M \end{array}  \right] ~=~
\left[ \begin{array}{cc} A & B \\ B & D \end{array}  \right] \,\cdot\,
\left[ \begin{array}{c} \epsilon \\ \kappa \end{array}  \right]\,,
\eeq
linearly relating mid-plane strains $\epsilon$ and plate curvatures $\kappa$ to vectors of running
loads $N$ and out-of-plane moments $M$. To simplify analysis we follow well-established industry
requirements according to which only the case of symmetric laminates with finite (usually small) number of
possible ply orientation angles $\theta_i \in \Theta = \{ \Theta_1, ..., \Theta_{N_\Theta}\}$,
are to be considered. Then $B$ tensor vanishes identically while in-plane and out-of-plane stiffness matrices
can be expressed as linear functions of four-dimensional lamination parameters $\xi^A_{i}$, $\xi^D_{i}$, $i = 1,2,3,4$
\beq
\label{xi-def}
\begin{array}{c}
\vec{\xi}^A ~=~ h^{-1} \, \sum^N_{i=1} \vec{\zeta}^{(0)}_i \ \Delta z \\
~\\
\vec{\xi}^D ~=~ 3 \, h^{-3} \, \sum^N_{i=1} \vec{\zeta}^{(0)}_i \ z^2_i \Delta z \\
~\\
\vec{\zeta}^{(0)}_i ~=~ [\,\, \cos(2\theta_i), \,\, \cos(4\theta_i), \,\,\sin(2\theta_i),\,\, \sin(4\theta_i)\,\,]^T\,.
\end{array}
\eeq
Here $N$ is the half of total number of plies each of thickness $\Delta z$,
with $h = N \Delta z $ being the panel half-thickness,
$\theta_i$ denotes orientation of $i$-th ply, $z_i$ stands for
the distance between mid-planes of laminate panel and $i$-th ply
and we used vector notations instead of explicit indices (this nomenclature is followed below).
It is apparent that in-plane mechanical loads are independent upon the stacking sequence
$\vec{\theta}$ and are the functions of lamination parameters $\vec{\xi}^A$ only,
while for out-of-plane constraints stacking sequence dependence is totally hidden in $\vec{\xi}^D$ vector.
Note that $\vec{\xi}^A$  lamination parameters are functions of only the number of plies $n_i$ laid up
at particular angle $\theta_i$ (ply numbers $\vec{n}$). Hence dependence upon $\vec{\xi}^A$ could equally be
written as $\vec{n}$ dependence, but reverse statement does not hold in general,
the function $\xi^A(n)$ is invertible only in special case $N_\Theta = 4$.
As far as buckling load constraint is concerned it brings essentially nothing new to the above picture. Indeed,
buckling load factor is linear in $D$ matrix coefficients (see, e.g., Ref.~\cite{Bettebghor})
and hence depends on stacking sequence only through the $\vec{\xi}^D$ lamination parameters.

Unfortunately, it is hardly possible to perform similar analysis for  ``non-universal'' constraints
mentioned above. These are extremely case-to-case dependent, include specific manufacturing requirements
for each particular product and thus are to be taken as most general stacking sequence dependent constraints.
Therefore, the composite weight optimization problem can be summarized as
\beq
\label{problem}
\begin{array}{c}
\min \, \sum_i \, n_i \\
~ \\
\begin{array}{cc}
\mbox{s.t.} & 
\left\{
\begin{array}{cl}
P( \vec{n} ) \, \ge \, 0 & \mbox{``universal''; ply number dependent} \\[2mm]
L( \vec{\xi}^D ) \, \ge \, 0 & \mbox{``universal''; lamination parameters dependent} \\[2mm]
S( \vec{\theta} ) \, \ge \, 0  & \mbox{``non-universal'';  stacking sequence dependent} \\[2mm]
\end{array}
\right.
\end{array}
\end{array}
\eeq

Suitable solution methodology crucially depends upon the problem size.
It is true that for thin laminates consisting of only a few plies
direct methods are applied well and outperform alternative approaches. However, in engineering
applications ply number $N$ for each plate might be of order one hundred, not mentioning
the need to consider many different plates glued together. Then direct treatment becomes extremely
inefficient and alternative methods need to be developed. Nowadays the most widely used approach is
to apply approximation techniques (see Refs.~\cite{approximation} for review), in which
``universal'' constraints $P$, $L$ are treated rigorously at outer level, while the last set of
``non-universal'' $S$-constraints are accounted for only approximately at inner stage,
iterating the whole procedure in case of large discrepancies between outer and inner levels.
The possibility of problem decomposition into outer/inner levels arises from crucial observation that none of
``universal'' constraints depend on stacking sequence directly. Thus at outer iteration we can formally
consider $\vec{\xi}^A$, $\vec{\xi}^D$ as independent variables and solve the auxiliary task\footnote{
Here we're slightly sloppy in notations and use $\vec{n}$ instead of $\vec{\xi}^A$ in view of above
mentioned correspondence between these quantities.}
\beq
\label{outer}
\min \, \sum_i \, n_i \qquad \mbox{s.t.~~} \left\{
\begin{array}{c}
P( \vec{n} ) \, \ge \, 0 \\[2mm]
L( \vec{\xi}^D ) \, \ge \, 0
\end{array}\right.
\eeq
to get ``optimal'' lamination parameters $\vec{\xi}^A_*$, $\vec{\xi}^D_*$, while at inner
iteration we have to ensure that $\vec{\xi}^A_*$, $\vec{\xi}^D_*$ are indeed realizable in term of
particular stacking sequence
\beq
\label{inner}
\begin{array}{c}
\min_\theta \, |\vec{\xi}^A_* ~-~\vec{\xi}^A(\vec{\theta})|^2 ~+~ |\vec{\xi}^D_* ~-~\vec{\xi}^D(\vec{\theta})|^2 \\[2mm]
\mbox{s.t.} \qquad S( \vec{\theta} ) \, \ge \, 0
\end{array}
\eeq

Successful examples of the above methodology include bi-level composite optimization and lamination
parameters approach \cite{toporov}, which differ only in formulation of outer level problem.
In bi-level treatment one assumes high homogeneity of laminated composite and then derives
simple proportionality of $A$ and $D$ matrices thus eliminating $\vec{\xi}^D$ parameters
from the list of design variables. Then the outer level problem becomes relatively easy to solve. 
However, these simplifications
are not coming for free: once stacking sequence dependence is abandoned ``optimal'' ply numbers
$\vec{n}_*$ are not obliged to respect $L$-type (e.g., buckling) constraints, which is the prime
disadvantage of bi-level optimization. Contrary to that, in lamination parameters approach one
keeps the stacking sequence dependence explicit at outer level and considers lamination
parameters $\vec{\xi}^A$, $\vec{\xi}^D$ as independent design variables.
It is crucial that no additional assumptions are introduced here, moreover, 
there is no need to check strain or buckling constraints as long as inner level optimization
matches the optimal $\vec{\xi}^A_*$, $\vec{\xi}^D_*$ values coming from outer iteration.
However, the acute problem is to define the feasible region of lamination parameters
(see Ref.~\cite{Bloomfield} and references therein) so that the inner problem might be solved successfully.
Unfortunately, there are only a few rigorous results on what is the feasible domain
of $\vec{\xi}^A$, $\vec{\xi}^D$ variables. Although it is known \cite{convex} that 
feasible domain of lamination parameters is convex, up to our knowledge no complete explicit
equations tightly bounding allowed $\vec{\xi}^A$, $\vec{\xi}^D$ values are available.

Our approach to composite materials optimization lies essentially in-between the above
bi-level and lamination parameters methods. Namely, we suggest to retain
both $D$-type lamination parameters $\vec{\xi}^D$ and ply numbers $\vec{n}$ explicitly
at outer optimization level, so that the formulation \eq{outer} remains valid verbosely.
As far as feasible region of $\vec{\xi}^D$ values is concerned, we show that at any fixed
ply numbers feasible domain $\Omega_\xi(\vec{n})$ of realizable lamination
parameters $\vec{\xi}^D$ is convex polyhedral body with $N_\Theta !$ vertices, each of which
corresponds to a particular ``extreme'' stacking sequences compatible of given ply numbers. Therefore,
our approach is essentially the equation \eq{outer} supplemented with feasibility requirement
\beq
\label{omega1}
\vec{\xi}^D \, \in \, \Omega_\xi(\vec{n})
\eeq
and explicit description of the feasible region $\Omega_\xi(\vec{n})$ as the convex hull
of $N_\Theta !$ points or equivalently as a set of linear constraints:
\beq
\label{omega2}
\Omega_\xi(\vec{n}) ~=~ \{ \vec{\xi} \,:\, A \xi^D \, \le \, b \}\,.
\eeq
\subsubsection*{Feasible Region of Lamination Parameters}

Derivation of the explicit form of lamination parameters feasible region $\Omega_\xi(\vec{n})$
at fixed ply numbers $\vec{n} = \{n_1,..., n_{N_\Theta} \}$ is similar in spirit to what had been
done in seminal paper \cite{Miki}. Namely, for given vector $\vec\lambda$ consider the problem
\beq
\label{poly}
\begin{array}{ccc}
~ & \vec{\theta}_* ~=~ \arg\max\limits_{\theta_i \in \Theta} \,\,\, (\vec{\xi}^D \cdot \vec{\lambda})\,,
& \qquad \Theta ~=~ \{ \Theta_1, ..., \Theta_{N_\Theta} \} \\[3mm]
\mbox{s.t.} &
\sum_{i=1}^N \, \delta( \theta_i - \Theta_k ) = n_k & ~ \\
\end{array}
\eeq
which is an elementary step of constructing outer polyhedral approximation to $\Omega_\xi(\vec{n})$, \cite{dubrovin}.
Here $\Theta_i$ denotes allowed ply orientation angles, $\delta(x)$ stands for Kronecker delta-function
and component-wise representation of $\vec{\xi}^D$ is given in \eq{xi-def}. We assert that
stacking sequences, which deliver the maximum to \eq{poly} for various inputs $\vec\lambda$, are such that plies
of the same orientation are adjacent
\beq
\label{sol}
\vec{\theta}^{\,(\mu)}_* ~=~
\{
\underbrace{\Theta_{k_1},...,\Theta_{k_1}}_{n_{k_1}},
\underbrace{\Theta_{k_2},...,\Theta_{k_2}}_{n_{k_2}},
\cdots
\underbrace{\Theta_{k_{N_\Theta}},...,\Theta_{k_{N_\Theta}}}_{n_{k_{N_\Theta}}} \}\,,\qquad
\mu = 1, ..., N_\Theta !
\eeq
Indeed, let us consider $i$-th term in the objective function \eq{poly}:
\beq
z^2 \cdot [\lambda_0 \, \cos(2\theta_i) ~+~ \lambda_1 \, \cos(4\theta_i) ~+~
\lambda_2 \, \sin(2\theta_i) ~+~ \lambda_3 \, \sin(4\theta_i)]
\eeq
Suppose that it is maximized at a particular ``best'' ply orientation angle $\Theta_k$. If there would be no
constraints in the problem, solution must have all plies aligned at the same angle as well
(these solutions constitute the vertices of famous Miki tetrahedron, Ref.~\cite{Miki}). In our
case imposed constraints limit the maximal allowed number of plies with ``best'' orientation and
actual task is to place these plies optimally in the stack. Due to the positivity of measure factor $z^2$,
which is maximal at composite skin layer and vanishes at laminate mid-plane,
it is evident that the preferable position for ply with $\Theta_k$ orientation is near the composite skin.
Repeating the same arguments for all $n_k$ plies laid-up at $\Theta_k$  we conclude that
they occupy continuous stack located near the boundary layer.
For the remaining ply orientations the same reasoning applies verbosely, the only
difference is that boundary layer is now located at smaller $z$. 

It follows immediately from the above that all realizable values of lamination parameters $\vec{\xi}^D$
are located within the convex hull of ``extreme'' stacking sequences \eq{sol}:
$\Omega_\xi(\vec{n}) \,\in\, \mathrm{Conv}( \vec{\theta}^{\,(\mu)}_* )$. 
Confronting this with known concavity of lamination parameters feasible domain \cite{convex}
we conclude that $\vec{\xi}^D$ feasible region at fixed ply numbers must coincide with
$\mathrm{Conv}( \vec{\theta}^{\,(\mu)}_* )$
\beq
\label{feasible_region}
\Omega_\xi(\vec{n}) ~=~ \mathrm{Conv}( \vec{\theta}^{\,(\mu)}_* )\,.
\eeq
This is the prime theoretical result of our paper, which opens the possibility to construct
efficient bi-level optimization scheme for laminated composite optimization problem \eq{problem}.
Indeed, Eq.~\eq{feasible_region} might be used to ensure that
outer level iterations always produce realizable optimal lamination parameters
so that no difficulties can arise at inner level. Moreover, the number of
vertices in corresponding convex polytope is relatively small and admits its efficient description
in terms of linear inequalities. Indeed, in the most practically important case we have $N_\Theta = 4$
and hence only $N_\Theta ! = 24$ vertices, for which convex hull construction rises no computational issues
whatsoever.

\subsubsection*{Acknowledgments}
We acknowledge thankfully fruitful discussions with S.M.~Morozov and S.~Lupuleac.


\end{document}